\documentclass[preprint,12pt]{aastex}

\newcommand\erfc{\mathop{\rm erfc}}

\newcommand\hdtwo{HD 209458}

\newcommand\hdtwob{HD 209458b}

\newcommand\spitzer{{\it Spitzer Space Telescope}}
\newcommand\selfname{paper} 
\newcommand\iraca{3.6~\micron{}}
\newcommand\iracb{4.5~\micron{}}
\newcommand\iracc{5.8~\micron{}}
\newcommand\iracd{8.0~\micron{}}

\slugcomment{\today}
\shorttitle{Resolving Planetary Surfaces with Light Curves}

\shortauthors{Williams et al.}

\begin{document}

\title{Resolving the Surfaces of Extrasolar Planets \\
With Secondary Eclipse Light Curves}

\author{Peter~K.~G.~Williams\altaffilmark{1}}
\author{David Charbonneau\altaffilmark{1}}
\author{Curtis~S.~Cooper\altaffilmark{2}}
\author{Adam~P.~Showman\altaffilmark{2}}
\author{Jonathan~J.~Fortney\altaffilmark{3}}

\altaffiltext{1}{Harvard-Smithsonian Center for Astrophysics, 60
  Garden Street, Cambridge, MA 02138; pwilliams@cfa.harvard.edu,
  dcharbonneau@cfa.harvard.edu}

\altaffiltext{2}{Department of Planetary Sciences and Lunar and
  Planetary Laboratory, University of Arizona, 1629 University
  Boulevard, Tucson, AZ 85721; curtis@lpl.arizona.edu,
  showman@lpl.arizona.edu}

\altaffiltext{3}{Space Science and Astrobiology Division, NASA Ames
  Research Center, MS 245-3, Moffett Field, CA 94035,
  jfortney@arc.nasa.gov}

\begin{abstract}

We present a method that employs the secondary eclipse light curves of
transiting extrasolar planets to probe the spatial variation of their
thermal emission. This technique permits an observer to resolve the
surface of the planet without the need to spatially resolve its
central star. We evaluate the feasibility of this technique for the
\hdtwo{} system by first calculating the secondary eclipse light
curves that would result from several representations of the planetary
emission, and then simulating the noise properties of observations of
this signal with the \spitzer{} Infrared Array Camera (IRAC).  We
consider two representations of the planetary thermal emission; a
simple model parameterized by a sinusoidal dependence on longitude and
latitude, as well as the results of a three-dimensional dynamical
simulation of the planetary atmosphere previously published by Cooper
\& Showman. We find that observations of the secondary eclipse light
curve are most sensitive to a longitudinal offset in the geometric and
photometric centroids of the hemisphere of the planet visible near
opposition.  To quantify this signal, we define a new parameter, the
``uniform time offset,'' which measures the time lag between the
observed secondary eclipse and that predicted by a planet with a
uniform surface flux distribution.  We compare the predicted amplitude
of this parameter for \hdtwo{} with the precision with which it could
be measured with IRAC.  We find that IRAC observations at \iraca{} of
a single secondary eclipse should permit sufficient precision to
confirm or reject the Cooper \& Showman model of the surface flux
distribution for this planet.  We quantify the signal-to-noise ratio
for this offset in the remaining IRAC bands (\iracb{}, \iracc{}, and
\iracd{}), and find that a modest improvement in photometric precision
(as might be realized through observations of several eclipse events)
should permit a similarly robust detection.

\end{abstract}


\section{Introduction}
\label{sec:intro}

The identification of the first transiting extrasolar planet \hdtwob{}
\citep{charb00,henry00,mazeh00} initiated a flurry of investigations 
into the properties of the planetary body that are not possible
for non-transiting objects.  Nine transiting extrasolar planets
have now been identified; for a review of their properties, see Charbonneau et al.\ 2006.  
In addition to HD~209458b, three of these (TrES-1; Alonso et al.\ 2004, 
HD~149026b; Sato et al.\ 2005, and HD~189733b; Bouchy et al.\ 2005) 
orbit stars that are sufficiently close and hence bright enough ($V < 12$)
to permit a direct study of their atmospheric absorption and emission features
through a variety of techniques.  One such method is that of 
transmission spectroscopy, whereby stellar spectra gathered outside and
inside times of planetary transit are compared to search for additional
absorption features in the latter due to the presence of certain
atoms or molecules in the planetary atmosphere.  The only detections
of this effect to date have been achieved with the STIS spectrograph aboard the 
\emph{Hubble Space Telescope}:  \citet{charb02} observed \hdtwo{} in visible light 
and detected the absorption from gaseous atomic sodium in the planetary atmosphere, 
and \citet{vidal03} observed the same system in the ultraviolet to detect
absorption resulting from a large cloud of atomic hydrogen 
escaping from the planet.  Numerous ground-based observational efforts 
(Bundy \& Marcy 2000; Moutou et al.\ 2001, 2003; Brown et al. 2002; 
Winn et al. 2004; Narita et al. 2005) have yielded only upper limits,
albeit useful ones.  Most recently, \citet{deming05a} placed a stringent
upper limit on the presence of gaseous CO from observations near 2.3~\micron. 
Along with the sodium detection, these observations place tight constraints on the 
distribution of condensates in the upper atmosphere (Fortney 2005).  

A complementary technique that promises to be at least as powerful is
that of occultation photometry and spectroscopy.  This method
subtracts observations gathered during secondary eclipse (i.e. when
the planet passes \emph{behind} the star) with those gathered just
before or after this time (when the planet is unocculted), to search
for any excess emission attributable to the planet itself.  For a hot
Jupiter planet orbiting a Sun-like star, the relative size of this
excess at infrared wavelengths is a few parts in one
thousand. Ground-based attempts to observe thermal radiation from
transiting planets have been frustrated by the variability of the
telluric opacity over short timescales, and the large, ambient thermal
background, and thus have resulted in only upper limits
\citep{wiedemann01,lucas02,richardson03a,richardson03b}.  Recently,
two groups have succeeded in detecting the planetary thermal emission,
and once again this feat was enabled by a space-based observatory, in
this case the \spitzer{} (Werner et al.\ 2004).  \citet{charb05} used
the Infrared Array Camera (IRAC; Fazio et al. 2004) to detect the
thermal emission of TrES-1 in two band passes, \iracb{} and \iracd{}.
\citet{deming05b} used the Multiband Imaging Photometer (MIPS; Reike
et al.\ 2004) to detect the thermal emission from \hdtwob{} at
24~\micron, and \citet{deming06} employed the Infrared Spectrograph
(IRS; Houck et al. 2004) to detect the emission from HD~189733b at
16~\micron.  Each of the data sets consisted of a rapid cadence, high
signal-to-noise ratio (SNR) photometric light curve spanning a
predicted time of secondary eclipse, in which a decrement in the total
system flux of the expected depth and duration was clearly detected at
the anticipated time.

An intriguing possibility permitted by such observations is
that of resolving the surface of the planet through high SNR
photometry during the times of ingress and egress, when different
portions of the planetary surface are occulted by the star.
The purpose of this paper is to explore this effect in detail,
and to evaluate the likelihood of its detection with the
\spitzer{}.

It is not currently possible to directly image
exoplanetary surfaces, nor is this ability anticipated for
any planned facility, including the NASA \emph{Terrestrial Planet Finder}.
Nonetheless, the spatial-dependence of the planetary photosphere is accessible
to an observer through careful monitoring of the structure in a 
secondary-eclipse light curve.  During the
ingress and egress phases of secondary eclipse, which last
approximately $15 - 30$ minutes depending on the orbital geometry and
planetary radius, the portion of the visible planetary hemisphere
(the ``dayside'') that remains unocculted varies smoothly as a function of time.
From the known system parameters (namely the orbital period, phase,
inclination, and radii of the planet and star), it is then possible to 
invert the observed light curve to recover some 
aspects of the flux distribution across the dayside of the planet.  This technique is not new; a
similar approach has been used, for example, to produce surface maps of Pluto and 
Charon (for a review for such observations, see Stern 1992). 
But only with the \emph{Spitzer} detections of the past year 
has it been feasible to consider applying this technique to extrasolar planets.

This possibility is particularly interesting because 
several recent dynamical simulations have predicted the presence of a  
large flux contrast across the dayside of a hot Jupiter planet.  
Independent studies of \hdtwob{} by \citet{showman02} and
\citet{cooper05} predict a strong eastward jet.  The
CS05 supersonic jet pushes the atmosphere's hottest region downstream
by 60 degrees from the planetary substellar point.  The planetary
circulation pattern leads to global temperature variations of
$\sim$500 K at photospheric pressures.  Simulations by \citet{cho03}
find three broad east-west jets and polar vortex motions.  They find
that temperature variations may reach up to $\sim$1000 K in certain
circumstances.  \citet{burkert05} find similarly large temperature
variations. They highlight how atmospheric opacity, which controls the
penetration depth of stellar flux, affects the circulation problem.
Rapid cadence light curves during ingress and egress may allow for
observational tests of these circulation models. The analysis of such
light curves would provide insight into the dynamical flows of hot
Jupiters and also illuminate the energy budgets of these planets.

In \S\ref{sec:motivation}, we further motivate the use of a secondary
eclipse light curve to study a planetary surface flux distribution. 
In \S\ref{sec:lightcurves}, we describe the software that
we developed for simulating light curves, which forms the basis of our
subsequent investigations. In \S\ref{sec:fits}, we introduce a 
parameter, the ``uniform time offset,'' which characterizes the
longitudinal flux contrast of a planetary emission. We then probe the
behavior of this parameter with Monte Carlo simulations of
hypothetical light curves. Finally, in \S\ref{sec:discussion},
we discuss our results and near-future applications of this
technique. In this \selfname, we restrict our focus
to upcoming observations of \hdtwob{} with
the \emph{Spitzer} IRAC instrument. It should be emphasized, however, 
that our technique can be applied to other instruments and
transiting extrasolar planets.

\section{Motivation}
\label{sec:motivation}

The key concept behind this \selfname{} is that the shape of
the ingress and egress portions of a secondary eclipse light curve
convey information about the spatial variation of the flux emitted
across the visible hemisphere of a planet. This is because the disk of the primary gradually and
predictably obscures the disk of the planet during secondary
eclipse. If the geometry of the orbit is known, at a given point in
time the observer knows exactly which region of the planet is
obscured. The time variation in the
light curve thus becomes a proxy for the spatial variation in emitted
flux over the unobscured region of the planetary photosphere.

For an observer, the issue is how one infers a flux distribution from
an observed light curve. It is not possible
to map uniquely from the latter to the former, as the
one-dimensional light curve can convey only an integrated version of
the planet's two-dimensional emission. Gross surface features,
however, can be inferred from a light curve. For instance, a relatively
large and rapid increase in flux during egress indicates that a bright spot 
on the planet was just unobscured. The location of this
spot on the planetary surface can be partially constrained based on the
known orbital geometry; for example, neglecting the curvature of the
stellar limb across the planetary surface, and approximating the orbital
inclination as 90\degr{}, the longitude of the
spot can be determined, but the latitude will remain unknown.
A noteworthy feature of secondary eclipse light curves is that their
ingress and egress portions provide complementary information about the
spatial variation of planetary flux, since 
ingress and egress permit distinct regions of the planet to be viewed
in isolation.  An effective way of extracting
information from a noisy eclipse light curve is to compare it to the output of a model; 
that is, to test whether a light curve is consistent with a given
underlying model of the planet surface flux distribution. 
This requires a well-defined measure of consistency that will be useful in the face of
realistic photometric noise that may partially corrupt the features of the ingress
and egress portions of a light curve.

An additional concern is that of planetary rotation.  In principle,
this method could be used to determine directly the rotation
period of the hot Jupiter, by observing the change in location
on the visible hemisphere of a feature during the elapsed time between ingress
and egress (typically $ 1.5 - 3$~hours, depending on the
orbital geometry and stellar radius).  Hot Jupiters are expected to
be tidally locked, however, in which case the planet rotates only slightly
during this time (specifically, \hdtwob{} rotates 13\degr{} during the eclipse).
Although we include the effect of such rotation in our calculations below,
we note that during ingress and egress a tidally-locked planet presents 
essentially the same face, thus simplifying interpretation of any observed
structure in the eclipse light curve.

\section{Modeling Light Curves}
\label{sec:lightcurves}

To simulate the eclipse light curve, we first calculated a 
series of resolved flux images of the planet as a function of orbital
phase.  Each flux image is then converted into a single flux
value by masking out the portion of the planet that would be obscured
by the primary and summing the unmasked pixels, and a complete
light curve is composed of a sequence of these flux values.
The flux values are then normalized to a predefined eclipse depth,
to match the expected planet-to-star flux ratio.

In all of our simulations, we fixed the orbital and system parameters
to those of the HD~209458 system (\S3.3).  Each light curve is a function of
the underlying model of the planetary emission, and the spectral response
function of the particular IRAC bandpass (\S3.2), which is required
to calculate the relative number of photons generated by each pixel.  
We consider three models for the spatial-dependence of the planetary emission, 
of increasing complexity: 
(1) a uniformly emitting disk, 
(2) a sinusoidal variation in longitude and latitude, and, 
(3) a numerical, dynamical simulation of the atmosphere of \hdtwob{}. 
For each of these 3 emission models, we generated four light curves, one for each bandpass of IRAC.
These individual steps are described in greater detail in the following sections.

\subsection{Flux Images}
\label{ssec:fluximages}

Each flux image depicts the planet as a disk projected onto a 256$\times$256
pixel grid; a greater resolution does not alter the resulting light curve,
as the underlying models for the spatial dependence of the flux do
not vary on finer spatial scales. For
the uniform emission model, every pixel in the disk is set to
unity. In the other cases, the value of each pixel in the disk is
derived from a flux map of the emission of the planet. The
coordinates of each pixel are reverse-mapped to a latitude $\phi$ and
longitude $\lambda$ using an inverse orthographic projection in which
the orientation of the planet is determined from the geometry of the
orbit and the assumption that the planet is tidally locked
with its primary. The point $(\phi = 0, \lambda = 0)$ is defined to be
the substellar point. In the sinusoidal model, the flux map
is specified as
\begin{equation}
F_{A,B}(\phi, \lambda) = A \cos(\phi) + B \sin(\lambda) + 1,
\end{equation}
where $A$ and $B$ are parameters that specify the latitudinal and
longitudinal contrast, respectively. 

\begin{figure}[htb]
\plotone{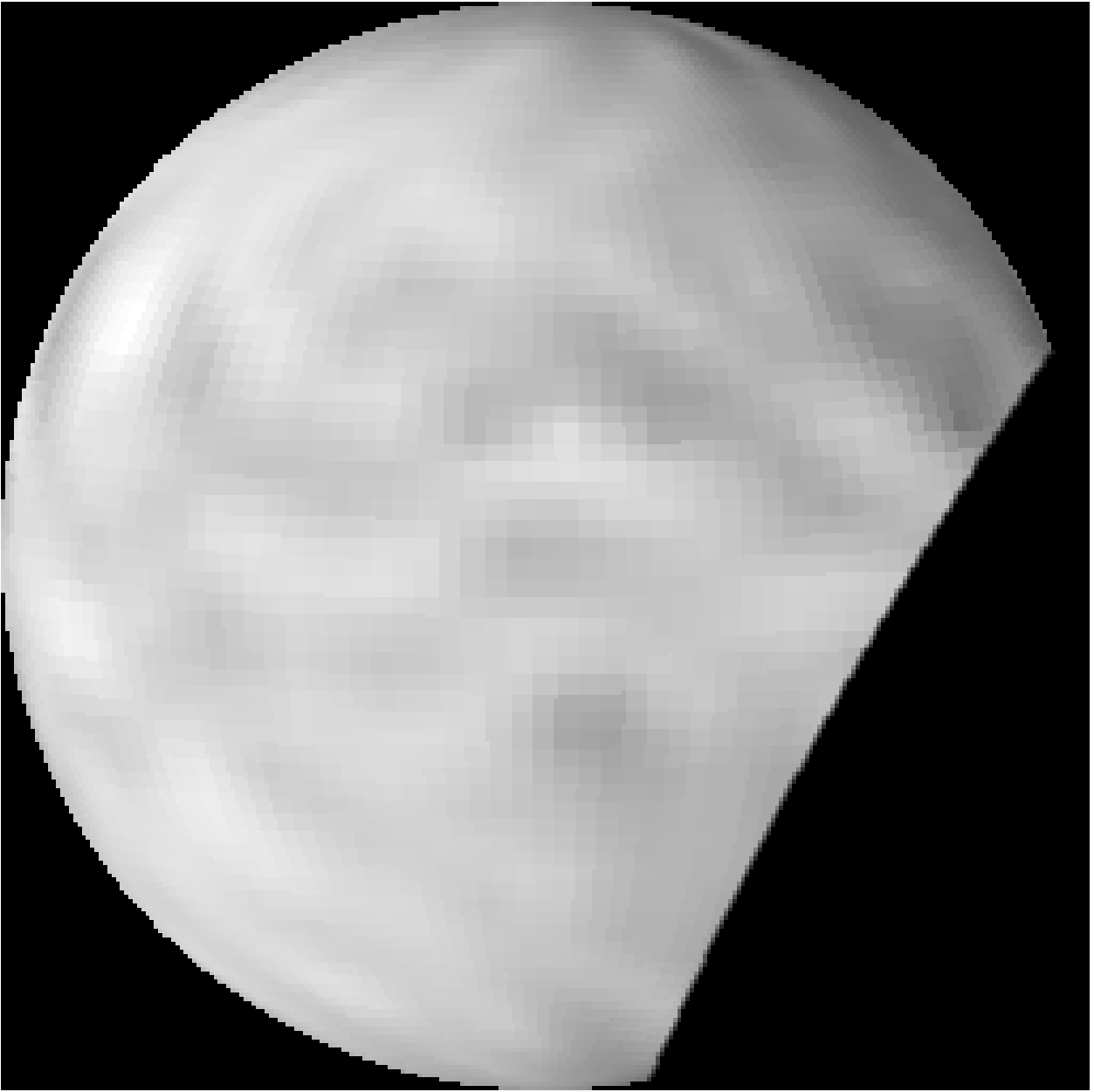}
\caption{Example flux image simulating the appearance of \hdtwob{} 
as described by the CS05 model in the \iracd{} IRAC bandpass
at a time five minutes after the start of ingress. 
The photospheric pressure for this bandpass is 24 mbar.} 
\label{fig:example_image}
\end{figure}

In the third case, we employ the results of the CS05 simulation,
which modeled \hdtwob's atmosphere 
as a gas in three dimensions with columns in hydrostatic equilibrium. 
Their model predicts a superrotating zonal jet dominating the flow at the
equator and mid-latitudes.  This jet blows the hottest regions of the
atmosphere downstream from the substellar point by about 60 degrees.
At the time of the secondary eclipse, this appears as a large, hot
region near the planet's trailing edge. 
CS05 ran the simulation with a
resolution of 45 points in latitude, 72 points in longitude, and 40
vertical layers logarithmically spaced from 1 mbar to 1 kbar.
The output of their model is the temperature and pressure at each
of these locations.
We interpolated the output of the CS05 model onto a grid in longitude
and latitude of twice their original resolution, and then
to each of our $256\times256$ points, we associated the nearest 
interpolated CS05 grid point, and assigned a flux value using the 
method described in \S3.2.  Figure~\ref{fig:example_image} shows an example flux
image derived from the CS05 model.

We also considered the simulation of \hdtwob{} developed by \cite{cho03}, who modeled the 
planetary atmosphere as a frictionless two-dimensional gas in hydrostatic balance
(in contrast to the three-dimensional model of CS05). 
Unlike the results for CS05 model, the Cho et al.\ model
resulted in only very weak perturbations to the secondary eclipse light curve.
Since such a light curve could not be distinguished from a uniformly
emitting disk (given foreseeable observational errors), we did
not pursue this model further.

\subsection{Temperature to Flux Conversion for the CS05 Model}
\label{ssec:temptoflux}

In order to generate flux images, the pressure-temperature (P-T)
profiles of the CS05 model must to be converted to flux values. 
Formally, this would require us to solve the equation
of radiative transfer in a dynamical, spatially-varying model of the 
chemistry of \hdtwob{}'s atmosphere.  Since our goal
in this paper is to present a first estimate of the perturbations
to the secondary eclipse light curve, we simplified the 
problem using the approach described below. We
model the emission of each atmospheric column as a blackbody emitting
isotropically at the temperature in the P-T profile corresponding to
the photospheric pressure. Photospheric pressures for various
wavelengths were previously determined from radiative transfer calculations by
\cite{fortneyet05}.  
Using the same one-dimensional HD209458b model, we computed the
brightness temperature in each of the IRAC band passes, taking into
account each band's transmission function\footnote{\tt
\url{http://ssc.spitzer.caltech.edu/irac/spectral\_response.html}}.
The brightness temperatures are different in each of the four bands
and are sensitive to atmospheric opacity.  We then compared these
brightness temperatures to the Fortney et al. pressure-temperature
profile to obtain the atmospheric pressure that corresponds to each
temperature.  One can think of each of these pressures as the
``photospheric pressure'' in each band.  Assuming solar composition and
approximate radiative equilibrium, we find resulting photospheric
pressures of 95 mbar (\iraca{}), 48 mbar (\iracb{}), 32 mbar
(\iracc{}), and 27 mbar (\iracd{}).  Four brightness temperature maps
were then generated by taking the temperatures from the pressure
levels in the CS05 tables nearest these photospheric pressures: 105
mbar, 50 mbar, 35 mbar, and 24 mbar, respectively. These maps were
converted to final flux maps by integrating the IRAC response
functions over blackbody emission at those temperatures.

This approach is approximate because the photospheric pressures were
calculated using \cite{fortneyet05}'s radiative-equilibrium P-T
profiles rather than the weather-modified P-T profiles from CS05.
Nevertheless, the temperature dependences of the opacities are modest,
so the photospheric pressures calculated in this manner should provide
reasonable approximations for the CS05 P-T profiles.  A significant
uncertainty is the existence of clouds: an opaque cloud at a few mbar
pressure, for example, would move the photosphere to the cloud-top
pressure in all four bands.  We have assumed that such clouds are
absent.  We note that significant opacity variations on isobars are
possible in principle, due to lateral variations in the concentrations
of CO and CH$_4$ in the upper layers of the planetary atmosphere.
Recent simulations by \cite{cooper06}, however, show that the
concentrations of CO and CH$_4$ are likely homogenized above the 1 bar
level. This lends credence to our adoption of a single pressure level
identifying the photosphere in each bandpass. We also note that our
emission model ignores effects due to the slant path for emission from
the limb of the planet, i.e.\ limb darkening.  Fortney et al.\ (2006b)
perform detailed radiative transfer calculations for the Cooper \&
Showman (2006) dynamical atmosphere model of \hdtwob{}.  Using these
same methods, we have analyzed the CS05 grid and find an effect
equivalent to a temperature change of -100 K at the planetary limb.
We attribute this small degree of limb darking to the fact that the
day-side atmospheric profiles of the CS05 and Cooper \& Showman (2006)
grids are nearly isothermal; isothermal atmospheres show no limb
darkening.  We discuss in \S\ref{sec:discussion} why this small effect
can safely be ignored.

\subsection{Light Curve Generation}

\begin{figure}[htb]
\plotone{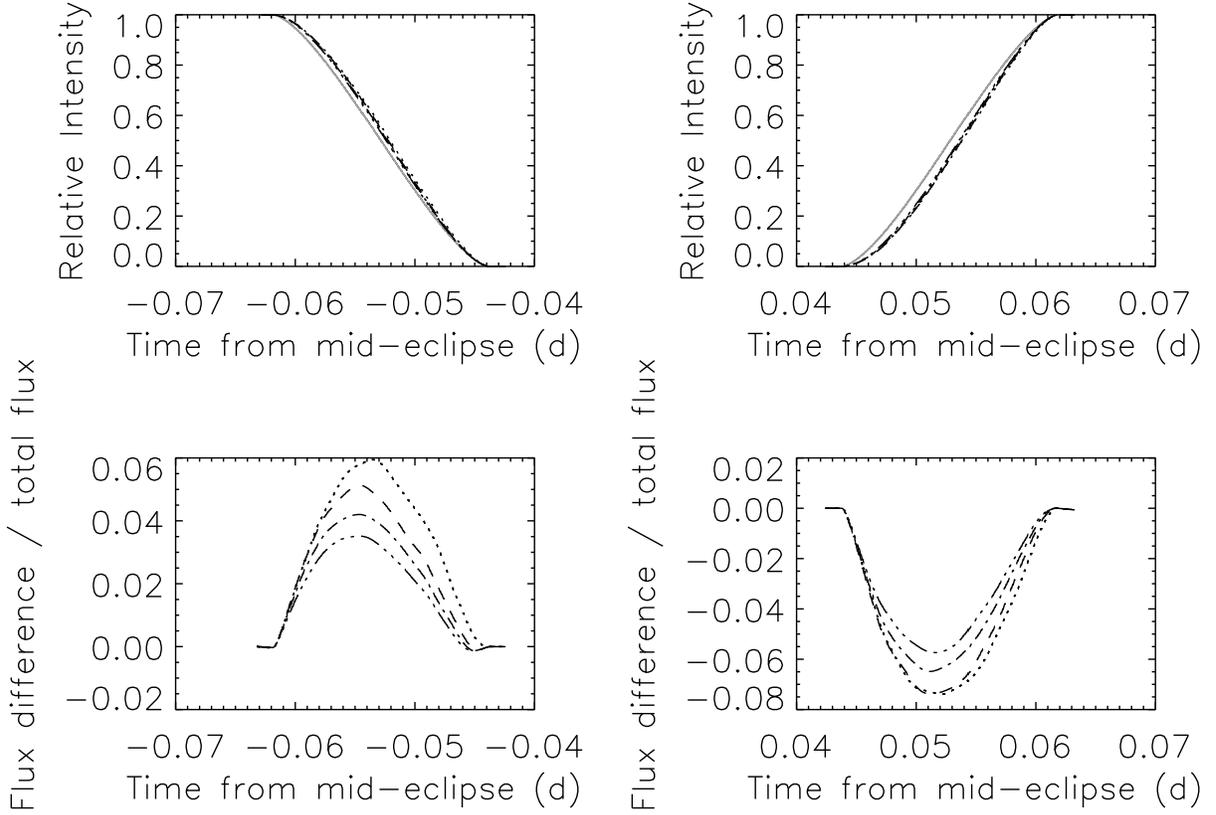}
\caption{\emph{Upper left panel:} Ingress portion of the secondary eclipse
light curves of \hdtwob{} assuming a uniform flux distribution (solid
gray line) and the CS05 model for the spatial flux variation (dashed lines overlying
each other). \emph{Lower left panel:} The relative differences in the curves resulting
from the CS05 model to the prediction of the uniform flux distribution, for
the four IRAC band passes (\iraca{}; dotted, \iracb{}; dashed, \iracc{}; dash-dotted,
and \iracd{}; dash-triple dotted).  The right panels depict these curves
and their differences at egress.}
\label{fig:cs05_curves}
\end{figure}

In each flux image, the portion of the planet obscured by the primary
is masked out to yield an image similar to
Figure~\ref{fig:example_image}. The position of the star relative to
the planet is calculated using the published system parameters: 
period $P$ = 3.52474 d, inclination $i$ = 86.6\degr, primary mass
$M_*$ = 1.1 M$_\sun$, stellar radius $R_*$ = 1.12 R$_\sun$, 
and planetary radius $R_P$ = 1.32 R$_J$. We assumed a circular orbit, 
consistent with the most recent radial-velocity data \citep{laughlin05} and
the timing of the secondary eclipse \citep{deming05b}. 
We represented the star as a geometric circle
with its center specified by the projected separation between the star and planet. 
Each pixel in the flux image was scaled by $n/4$, where $n$
was the number of the pixel's corners not inside the circle. 
The resulting light curve was then renormalized to account
for the flux ratio of the star to the planet. 
The band-dependent
depth of the eclipse was taken from theoretical calculations. 
\citet{fortneyet05} calculated the spectrum for HD~209458b from a radiative-equilibrium 
atmosphere model, and subsequently estimated the planet-to-star flux ratios
in each IRAC bandpass. 
They predict secondary eclipse depths of 0.00096 (\iraca{}), 0.00112 (\iracb{}), 
0.00145 (\iracc{}), and 0.00190 (\iracd{}), and we assume these
values here. Our resulting light curves for \hdtwob{} are
shown in Figure~\ref{fig:cs05_curves}.

We note here a few additional concerns.
First, we have neglected variations in the stellar flux 
over the $\sim$6~hour timescales, an assumption which is justified
by solar observations (see Batalha et al.\ 2002 and Borucki et al.\ 2004). 
We also neglect gravitational lensing by the primary, as calculations by
\cite{agol02} indicate that this effect is negligible. Finally,
the light curves are sensitive to the chemistry and dynamics of the
planetary atmosphere. The characteristics of both these aspects of the
atmosphere and the interplay between them (e.g.~clouds) remain
uncertain, but will be further constrained by both future
models that self-consistently couple dynamics and radiation, as well as
additional \emph{Spitzer} observations.

\section{Evaluating the Detectability of Surface Non-Uniformities}
\label{sec:fits}

The effect of a non-uniform flux distribution on 
the secondary eclipse light curves (Figure~\ref{fig:cs05_curves}) is subtle.
The detailed structure of ingress and egress is likely beyond
the reach of current instrumentation.  Nonetheless, it may
be possible to confirm or exclude the presence of large scale non-uniformities 
such as those predicted by the CS05 model.  To this end, we define
(\S4.1) a robust observational parameter that is sensitive to
such non-uniformities, and subsequently, using Monte Carlo
simulations (\S4.2), we evaluate its sensitivity for simulated 
\emph{Spitzer} IRAC data (\S4.3 \& \S4.4).

\subsection{The Uniform Time Offset}
\label{ssec:uto}

We define the ``uniform time offset'' $t_{offs}$ as the time
lag that must be applied to a synthetic light curve (generated
under the assumption of a uniform planetary flux distribution)
to minimize its ${\chi}^2$ difference from the data.
This parameter can be interpreted as the longitudinal separation 
between the photometric and geometric centroids of the planet,
i.e. the greater this separation, the greater the time lag 
between the center of the observed light curve (which is a function of the 
photometric centroid) and that of the predicted light curve (which is a 
function of the planet's projected location and hence geometric centroid). 
To first order, a planet's photometric centroid will be
off-center if its emission is longitudinally asymmetric, so the
uniform time offset is a simple observational gauge of this
longitudinal asymmetry. In detail, the time lag will also depend
on the orbital inclination and the curvature of the stellar limb
across the face of the planet.  We explore the behavior of this
parameter below.

To estimate the uniform time offset for a given observed time series, 
we must first generate a synthetic light curve. 
This uniform light curve $\Phi(t)$ is calculated under the assumption that
the planet has a uniform surface flux distribution,
and hence its tabulation requires only a knowledge of the orbital
parameters and the radii of the planet and star.
The eclipse depth is then normalized to match the estimated depth
of the observed time series.  Synthetic data points only need to be
generated for the ingress and egress portions of the transit, 
since the assumed uniform-flux model reveals no structure outside of these
times.  
In our implementation, we generated $\Phi(t)$
on a grid of times from 30~s prior to ingress to 30~s after egress;
this was smoothed with a boxcar function to eliminate any discrete pixel
effects resulted from our adoption of a $256\times256$ grid, and
subsequently interpolated to a spacing of 0.1~s.  
Once the synthetic light curve has been generated, the uniform time
offset can be evaluated by minimizing the $\chi^2$ value of the
model $\Phi(t)$ to the observed time series,
\begin{equation}
\chi^2(\tau) = \frac{1}{\sigma_{obs}^2} \sum_i{[F_i - \Phi(t_i + \tau)]^2},
\end{equation}
where $\tau$ is the time shift being applied, $F_i$ is the observed relative
flux value of the $i^\mathrm{th}$ data point, and $t_i$ is the time associated with 
that point. In our implementation, every point
in the observed light curve is given equal weight. 
We note that a possible variation would be to
determining distinct uniform time offsets for the ingress
and egress portions of a light curve, which would allow the observer
to investigate any differences between these two events.

To illustrate how the uniform-time offset encodes information about
the flux distribution, consider the synthetic, noiseless curves 
shown in Figure~\ref{fig:cs05_curves}.  The uniform time offsets 
for each of these are -86~s (\iraca{}), -77~s (\iracb{}), -67~s (\iracc{}), 
and -57~s (\iracd{}). The negative values imply that the photometric centroid of the 
planet lags the geometric centroid (and hence the expectations of a 
uniform flux planet).  Specifically, the planet exhibits 
an IR-bright region on the trailing limb, which is occulted in the later half
of the ingress and unocculted in the later half of the egress.  
The variation in the amplitude of the timing offset with bandpass encodes
critical information about the advection of heat and the strength of
winds as a function of depth in the atmosphere.  
In the CS05 model, the uppermost layers of the atmosphere
have a radiative equilibrium time constant that is much shorter than the
timescale over which winds act to advect heat across a hemisphere;
as a result, the hottest point is coincident with the subsolar point.
At greater atmospheric pressure (i.e. deeper in the atmosphere), this timescale
increases, and the hottest region is shifted significantly downwind (lags) 
the subsolar point.  Since decreasing wavelength corresponds to increasing photospheric 
depth (\S3.2), the bluemost IRAC bands are predicted to show the greatest value for
the uniform time offset.

\subsection{Monte Carlo Simulations with Synthetic Observations}
\label{ssec:montecarlo}

We generated synthetic IRAC observations of \hdtwob{} to
explore the detectability of the uniform time offset parameter. We chose
this facility because it is the optimal observatory for such observations,
and we can ground our calculations in the practical experience of
\cite{charb05}.  We created synthetic light curves spanning 0.1 days
from the center of eclipse, with a cadence of 15~s. To each model value,
we added Gaussian noise as estimated from the \emph{Spitzer} Observer's 
Manual\footnote{\tt\url{http://ssc.spitzer.caltech.edu/documents/som/}}.
These predictions are dominated by the photon noise from
the star, with a relative noise amplitude ${\sigma}_{obs}$ of 0.000323 (\iraca{}), 
0.000422 (\iracb{}), 0.001041 (\iracc{}), and 0.000831 (\iracd{}). 
We subsequently estimated the best-fit value of
the uniform time offset $t_{offs}$ by minimizing Eq.~(2).
%
We then conduct a Monte Carlo simulation by repeating this procedure
several thousand times, and creating a histogram of the derived values of
$t_{offs}$.  The resulting histogram is well-approximated by a
Gaussian, because the uniform time offsets derive from a simple fit to
a large number of data points, the noise properties of 
which are described by a Gaussian distribution.  

The resulting histogram then permits us to directly evaluate the
likelihood of detecting a non-zero value for $t_{offs}$ for a given
underlying model for the surface flux distribution.  We fit a Gaussian
to the histogram to estimate its mean $m$ and standard deviation
$\sigma$.  If a uniform time offset of $t_{offs}$ is calculated for
some set of observations, the probability $P$ of having obtained such
a value if the underlying model is correct decreases as $|m -
t_{offs}| / {\sigma}$.  We define $P$ to be the probability of making
an observation at least as far away from the expected mean as the
observed $t_{offs}$, which is given by the error function, $\erfc(|m -
t_{offs}| / \sqrt{2} \sigma)$.  We then consider this quantity for
$t_{offs} = 0$ (which we denote $P_{0}$) associated with a given model
and set of parameters.  $P_0$ is the likelihood that we would observe
a $t_{offs}$ of zero if the underlying model is correct. If $P_0$ is
small enough, then an observed $t_{offs}$ of zero would permit us to
rule out that model under consideration.  Conversely, if we find that
$P_0$ is not sufficiently small, then an observed $t_{offs}$ of zero
cannot rule out the model. An observed $t_{offs}$ large compared to
$\sigma$, however, would be inconsistent with the model.

\subsection{Sinusoidal Model Simulations}
\label{ssec:sinusoidal}

\begin{figure}[htb]
\plotone{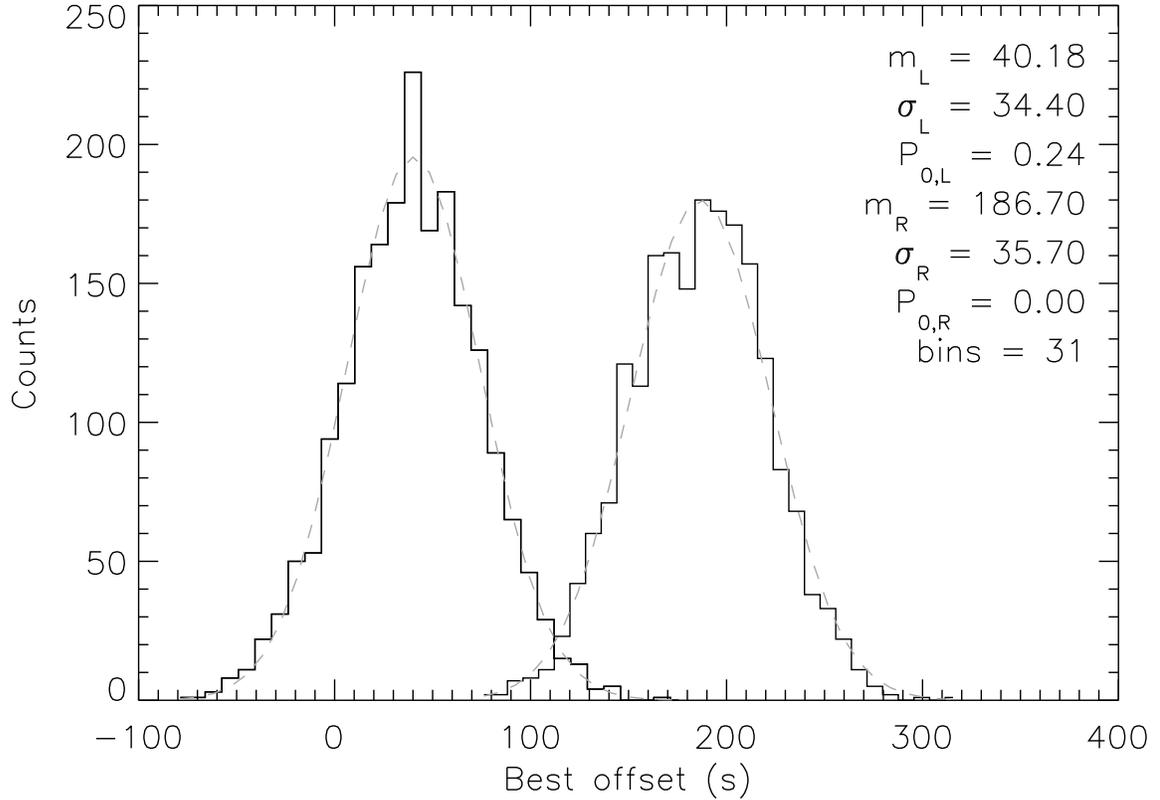}
\caption{Histogram of 2000 best-fit values of the
uniform time offset $t_{offs}$ for two pairs of parameters describing the
sinusoidal model, $(A,B) = (0.20, 0.20)$ \emph{(left solid line)} and 
$(A,B) = (0.10, 0.80)$ \emph{(right solid line)}.  
The observations are simulated in the \iracb{} IRAC band. 
The dashed curves are Gaussian fits to the
histograms, from which the mean $m_X$ and error $\sigma_X$ were
derived, with $X = L$ for the left curve and $X = R$ for the right
curve. The probability value $P_{0,L}$ for the left curve
is 0.24, and hence an observed value of $t_{offs} = 0$ cannot rule
out this model.  For the right curve, $P_{0,R}$ is essentially zero,
and thus an observed value of $t_{offs} = 0$ would exclude this
model with a high degree of confidence.}
\label{fig:sinemmc_2218}
\end{figure}

\begin{figure}[htb]
\plotone{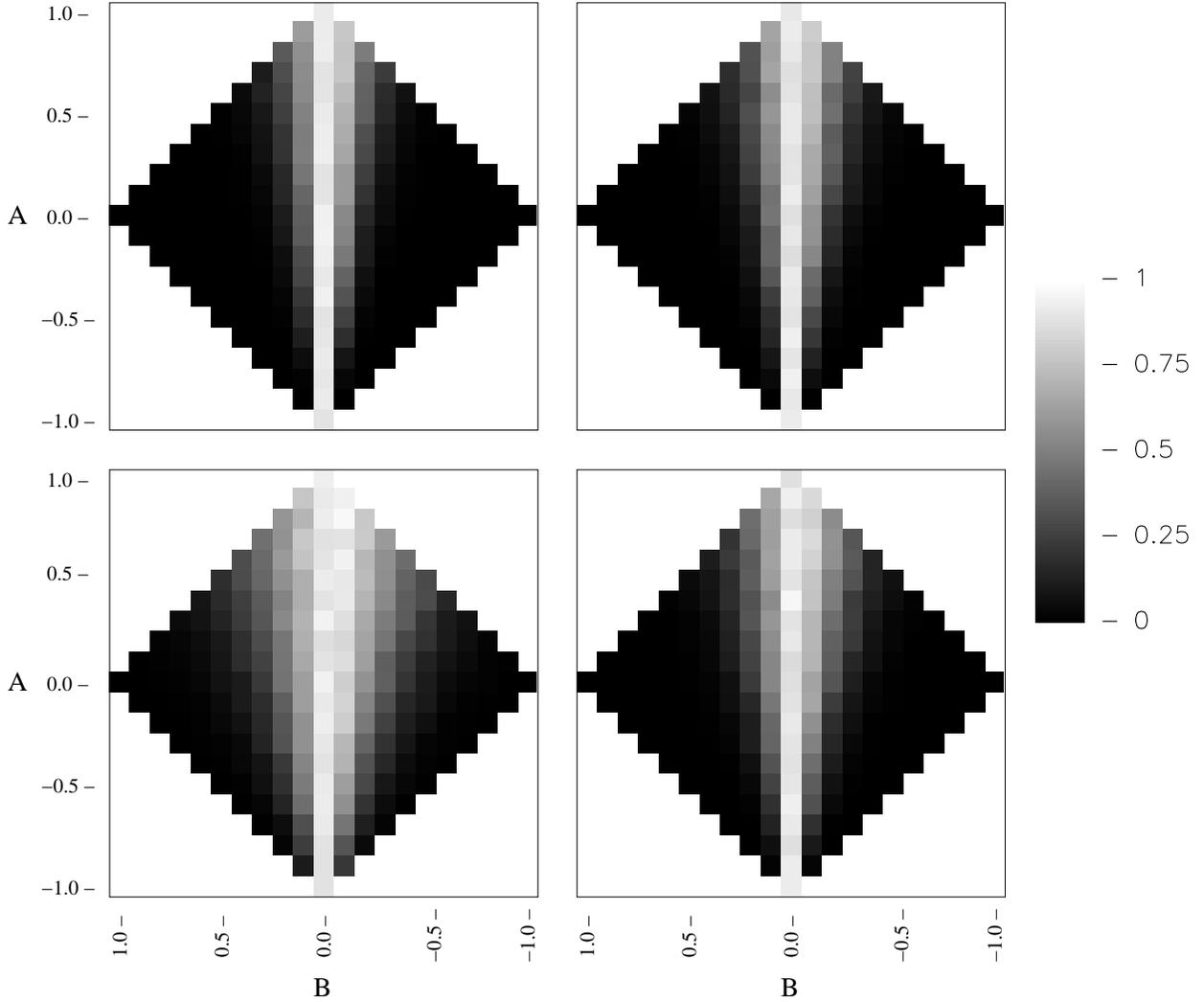}
\caption{The diamonds represent $P_0$ values (\S4.2) over the $(A,B)$
parameter space of the sinusoidal planetary model as determined by
Monte Carlo simulations of IRAC observations 
(\emph{upper left,} \iraca{}; \emph{upper right,} \iracb{}; 
\emph{lower left,} \iracc{}; and \emph{lower right,} \iracd{}). 
If a value of $t_{offs} = 0$ were to be observed for the \hdtwo{} system, 
models corresponding to the dark squares could be excluded with high confidence,
whereas models corresponding to the lighter squares could not.
The $A$ parameter controls the latitudinal flux contrast while the $B$ parameter
controls the longitudinal flux contrast.  
Although the dominant sensitivity is to longitudinal contrast,
note the effect of the non-equatorial orbit, which permits
a modest sensitivity to latitudinal flux variations.
Models with $|A| + |B| > 1$ were not considered because they would yield areas of
the planet presenting a negative flux.}
\label{fig:sinemmc_hd209}
\end{figure}

We ran a set of simulations with data generated from the sinusoidal
model described in \S\ref{ssec:fluximages}.
We systematically varied the $A$ and $B$ parameters (and hence
the degree of latitudinal and longitudinal contrast)
to find the subset of models which could reasonably be detected
with \emph{Spitzer} IRAC observations of \hdtwo{}.
We varied each of $A$ and $B$ from -1 to 1 with increments of 0.1,
considering all pairs for which $|A| + |B| < 1$ (values outside
this range would result in negative flux values at various points on the planet).
For each $(A,B)$ pair, 2000 simulated data sets were generated,
and from each we derived the value of $t_{offs}$. 
Figure~\ref{fig:sinemmc_2218} shows histograms for two
examples, $(A,B)$ = $(0.2,0.2)$ and $(0.1,0.8)$, both
modeled for the \iracb{} band. The former model is not easily
distinguished from one resulting from a uniformly emitting planet,
but the latter model has a strong longitudinal contrast that
should be readily detectable.  Figure~\ref{fig:sinemmc_hd209}
summarizes the detectability of given pairs of $(A,B)$ for the 4 IRAC
bandpasses.  Indeed, if the planet presents a large longitudinal contrast,
we are very unlikely to estimate a zero uniform time offset, while those with low contrast are expected
to yield offsets indistinguishable from zero. The latitudinal contrast has only a weak
effect, but interestingly it is not zero:  This results from the fact that the orbital inclination
of \hdtwob{} is not 90\degr. As a result, the projected limb of the star
across the planet is not symmetric in latitude, and hence the
ingress and egress curves encode some information of the latitudinal
flux distribution.

\subsection{Observability of CS05 Model}
\label{ssec:improvement}

\begin{figure}[htb]
\plotone{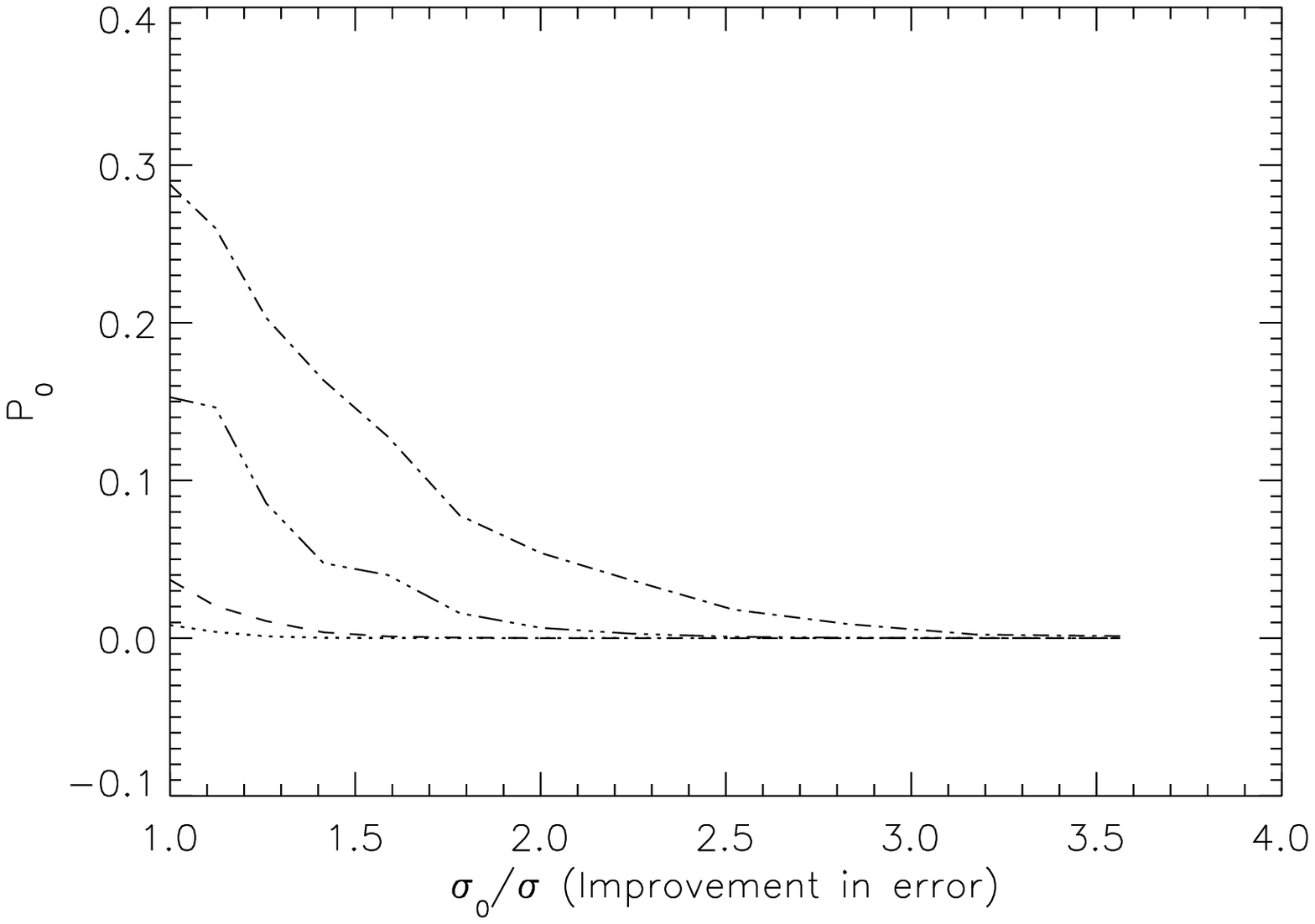}
\caption{Probabilities $P_0$ of observing a uniform time offset of 0
if \hdtwob{} is represented by the CS05 model, as a function of the
size of the observational errors relative to their nominal
\emph{Spitzer} IRAC values, ${\sigma}_0$. The four curves correspond to
the different IRAC band passes:  dotted; \iraca{}, dashed; \iracb{}, 
dash-dotted; \iracc{}, and dash-triple-dotted; \iracd{}.  We find that
IRAC observations at \iraca{} should provide a robust evaluation of the
predictions of the CS05 model.}
\label{fig:cs05_probs}
\end{figure}

We also ran a set of simulations to evaluate whether the CS05 planetary model discussed in
\S\ref{ssec:fluximages} could be tested with foreseeable \emph{Spitzer}
IRAC observations. In these simulations, we varied the photometric
precision ${\sigma}_{obs}$ and evaluated the resulting values for $P_0$. These
results then permit us to understand to what degree we must
improve the photometric precision of such observations to test
the CS05 model.
We initially used values of $\sigma_{obs}$ as currently estimated
for planned IRAC observations of \hdtwo{} as described in \S\ref{ssec:montecarlo}, 
and then decreased the assumed noise by factors of $4^{1/12}$. (We
conducted 12 runs, with a net improvement of a factor of 4 
in the overall precision.)  For each IRAC band and value of $\sigma_{obs}$, a histogram of 2500
uniform time offsets was generated, from which we calculated the value
of $P_0$. In Figure~\ref{fig:cs05_probs}, we show how $P_0$ decreases 
as a function of the assumed value for $\sigma_{obs}$ (relative to the
nominal value we expect IRAC to deliver) for each of the four IRAC
bands. In Table~\ref{tbl:cs05_curstats}, we list the values of
$m$ and $\sigma$ describing the resulting histogram of values of $t_{offs}$
derived for the CS05 model with the nominal IRAC errors.

We found that an observation of a zero uniform time offset in the
\iraca{} IRAC bandpass would already put the CS05 model into doubt,
since this would be expected to occur by chance in less than 1\%
of such data sets.  In the other IRAC bands, errors would need to be improved 
by factors of roughly 1.3 (\iracb{}), 3 (\iracc{}), and 2 (\iracd{})
to achieve a similar level of confidence.  Combining the statistical
significance of all four bands would permit the observer to
confidently exclude the CS05 model should a value of $t_{offs}$ of 0
be observed. 
We conclude that the uniform time offset technique is a useful
tool for testing the predictions of models similar to CS05,
owing to the prominent longitudinal flux contrast that they
present.  

\section{Discussion and Conclusions}
\label{sec:discussion}

In this \selfname, we have explored how the secondary eclipse
light curves of an extrasolar planet may be used to learn about the
spatial variation of its emission. This technique permits an observer
to gain access to this important information without the need to directly
image to surface of the planet. We have explored this technique with simulations of 
\emph {Spitzer} IRAC observations of the planet \hdtwob{} and considered the 
resulting value of the uniform time offset parameter $t_{offs}$
that would be observed if the emission of the
planet is consistent with a family of simple models whereby the
flux varies sinusoidally in longitude and latitude, as well
as the results of a three-dimensional dynamical calculation of the
atmosphere of this planet as previously published by Cooper \& Showman (2005).

We found that physically reasonable longitudinal flux contrasts could
plausibly be detected in the \hdtwo{} system with planned
\emph {Spitzer} IRAC observations. Specifically, sinusoidal models with 
a latitudinal parameter $|B| > 0.5$ yield a value of 
$P_0 < 0.25$ in all but the 5.6~\micron{} bands. From the definition
of the sinusoidal model, we see that $B = 0.5$ corresponds to a flux
contrast of a factor of 3 between the leading and trailing edges of
the planet. In the IRAC \iracb{} band this contrast
would correspond, for example, to two regions emitting as blackbodies with temperatures
of approximately 1300K and 900K. We find that the results of the
CS05 model are readily testable with \emph{Spitzer} IRAC observations.
In particular, observations in the \iraca{} band would already be able
to exclude this model with a high degree of confidence, should a value of $t_{offs} = 0$ 
be observed.  Of course, interpretation of such data will undoubtedly
be more complex that the approach presented here.  For example,
the presence of high-altitude clouds could decrease the atmospheric
pressure corresponding to the photosphere, which would serve to
mask the underlying dynamics of the atmosphere \citep{showman05},
where the effect of winds on redistributing the energy of the incident
stellar flux is much more prominent.

It should be noted that the value of the orbital eccentricity $e$ and
longitude of periastron $\omega$ also serve to affect the time of
secondary eclipse, which could mimic the effect discussed here.
Tidal circularization is expected to reduce the eccentricity of hot Jupiter
orbits to virtually zero, in which case observations of the primary
eclipse (e.g.  \citet{brown01}) of such systems can generally
constrain the predicted time of secondary eclipse to several seconds,
which contributes a negligible source of error to the value of
$t_{offs}$. Direct observational constraints on the orbital eccentricity
at the required level of precision are likely not feasible, however.
The current upper limit on the orbital eccentricity of HD~209458
from the radial velocity observations alone is $e<0.02$
(\citet{laughlin05}), which could induce an offset in the time
of secondary eclipse as large as 65 minutes, more than two orders
of magnitude greater than the effect we describe.
Rather, the validity of the assumption of a circularized orbit
can be bolstered by checking for the presence of transit timing
variations \citep{agol05,holman05} using a sequence of transit observations.
The observed lack of such variations (Knutson et al. 2006)
indicate the absence of perturbers in the \hdtwo{} system,
thus permitting tidal effects to complete the circularization of the
orbit.  We also note that the timing offset due to a non-zero orbital
eccentricity is fundamentally achromatic, in contrast to the
signal we describe here (e.g. Table~1).  By observing a single
secondary eclipse simultaneously in multiple bandpasses (such as the
4 bands of the IRAC instrument), an observer can search for variations
in the observed value of $t_{offs}$ between the various wavelength
bands, a sure sign that the signal is not due to residual orbital
eccentricity.  In particular, comparison between the values of $t_{offs}$
between the 3.6~$\mu$m and 8.0~$\mu$m bands would show the largest
effect.

There are several ways in which the predictions presented here can be
refined. More precise values for the orbital parameters and an
observational determination of the eclipse depths could refine our
theoretical predictions. But the greatest source of uncertainty in our
results is the model of the planetary atmosphere and its emission. As
mentioned before, clouds might dramatically affect the simulation
results. As alluded to in \S\ref{ssec:temptoflux}, limb darkening is
another effect that we do not take into account. Planets other than
\hdtwob{} may even exhibit limb brightening: this possibility has been
investigated with regards to the planet HD 149026b by
\cite{fortney06a}. Depending on their assumed model parameters, they
produce physically-plausible atmospheres that exhibit either limb
brightening or limb darkening. In either case, however, the phenomenon
is radially-symmetric and does not significantly alter the longitudinal flux
contrast of the planet, and so the effect on $t_{offs}$ should be
modest. We have confirmed this intuition by performing Monte Carlo
simulations with the CS05 model modified to have its emission scaled
according to the classical limb-darkening law $I'(r) = I(r) * (1 - c[1 -
\sqrt{1 - r^2}])$, where $r \in [0, 1]$ is the normalized projected
radial distance of a point from the center of the planetary disk and
$c$ sets the magnitude of the effect. With $c = 0.25$, equivalent to a
temperature decrement of $\sim$300~K on the limb of the planet (which
is much larger than what is calculated for \hdtwob{}), the resulting
change in $t_{offs}$ for all IRAC bands is $\approx3$ s.

In the near future, there will be many opportunities
to employ this technique for probing exoplanets. This \selfname{} has
considered the specific case of observations of \hdtwob{}
with the \spitzer's IRAC. Such observations should yield unprecedentedly
precise secondary eclipse light curves, and we conclude that
they can realistically  be expected to probe the surface flux 
distribution of the day side of the planet. They would
also determine the secondary eclipse depths of \hdtwob{} in the 4
IRAC bands, which would be of immediate interest in their own right. 
Even more promising would be observations of the recently-discovered 
extrasolar planet HD 189733b \citep{bouchy05},
which is a mere 19 pc away and has a very favorable planet-to-star flux
ratio, owing to the relatively large planet-to-star surface area ratio,
and the large equilibrium temperature of the planet. 
\emph{Spitzer} has already observed HD~189733 with all 3 instruments,
and we encourage a search for the effects described in this paper. 
In the longer term, the {\em James Webb Space Telescope} (JWST),
with infrared detectors and a larger aperture than {\em Spitzer}, will be
an extremely powerful tool for applying this technique. As
instrumentation and models improve, and as an increasing number of 
nearby transiting extrasolar planet systems are discovered, the
prospects for resolving the surfaces of these distant worlds grows
ever brighter.

\acknowledgments

We would like to thank J.~Y-K.~Cho, K.~Menou, B.~M.~S.~Hansen, and
S.~Seager for providing the results from their simulations for use in
light curve simulations. We thank the anonymous referee for comments
that improved the manuscript.  This work is based in part on
observations made with the Spitzer Space Telescope, which is operated
by the Jet Propulsion Laboratory, California Institute of Technology
under a contract with NASA. Support for this work was provided by NASA
through an award issued by JPL/Caltech.  The contributions of
C. Cooper and A. Showman were supported by NSF grant AST-0307664 and
NASA GSRP NGT5-50462. J.~Fortney is supported by a National Research
Council Fellowship.

\clearpage

\begin{table}[htb]
\begin{center}

\caption{Parameters for the histogram of uniform time offsets of the CS05 model
  \label{tbl:cs05_curstats}}

\begin{tabular}{lrrrr}
 & \iraca{} & \iracb{} & \iracc{} & \iracd{} \cr
\tableline
\tableline
observational precision, $\sigma_{obs}$ & 0.000323 & 0.000422 & 0.001041 & 0.000831 \cr
\tableline
mean, $m$ (s) & -91.44 & -80.01 & -76.91 & -62.61 \cr
standard deviation, $\sigma$ (s) & 34.63 & 38.35 & 72.34 & 43.79 \cr
\end{tabular}
\end{center}
\end{table}

\clearpage

\end{document}